\documentclass[notitlepage,nofootinbib,floatfix,amsmath,amssymb,showpacs,showkeys,superscriptaddress,prd,10pt]{revtex4-1}
\usepackage{graphicx,epstopdf}
\usepackage{subfig}
\def\qslash{q\!\!\!\slash }
\def\xslash{x\!\!\!\slash }

\begin{document}

\title{Isovector axial vector and pseudoscalar transition form factors of $\Delta $  in QCD}

\author{A. Kucukarslan}%
\affiliation{Physics Department, Canakkale  Onsekiz Mart University, 17100 Canakkale, Turkey}
\email{akucukarslan@comu.edu.tr}
\author{U. Ozdem}%
\affiliation{Physics Department, Canakkale  Onsekiz Mart University, 17100 Canakkale, Turkey}
\author{A. Ozpineci}%
\affiliation{Physics Department, Middle East Technical University, 06531 Ankara, Turkey}
\email{ozpineci@metu.edu.tr}

\begin{abstract}
We investigate the isovector axial vector and pseudoscalar form factors of $\Delta$ baryon by employing
light-cone QCD sum rules. Numerical calculations show that the form factors can be well fitted by the
exponential form. We make a comparison with the predictions of lattice QCD, chiral perturbation theory and quark model.
\end{abstract}
\date{\today}
\maketitle
\section{Introduction}

Form factors are important properties of hadrons.
They describe how hadrons interact with each other and give information about the internal structure of the hadrons such
as the measure of the charge and current distributions.
The $\Delta(1232)$ resonance is the lightest excitation of the nucleon.
Solving the mysterious structure of the $\Delta$ resonance has a significant effect to nuclear phenomenology.
The $\Delta$ is a rather broad resonance close to the $\pi N $ threshold. Studying the structure of the  $\Delta$
resonance theoretically of using lattice QCD's  an important input to phenomenology that cannot be directly extracted from
experiments. This is because the $\Delta$ decays strongly with a lifetime of  $10^{- 23}$seconds
\cite{LopezCastro:2000cv, Kotulla:2002cg} and resists experimental probes. Measurements of the $\Delta$ magnetic moment exist despite
a large experimental uncertainty. Moreover, the $\Delta$ was studied within the framework of the  lattice \cite{Alexandrou:2013opa}.
The decomposition of the $\Delta(1232)$ matrix elements into the appropriate Lorentz invariant form factors was carried out
and techniques to calculate form factors were developed and tested using quenched configurations. Besides, the axial charge $g_A$ is an important parameter for low-energy effective theories. $ g_A$ can also be viewed as an indicator of the
phenomenon of spontaneous breaking of chiral symmetry of nonperturbative QCD \cite{Choi:2010ty}.

Form factors are nonperturbative objects. Therefore, in order to study form factors, one needs to  use a non-perturbative method. QCD sum rules is one of the nonperturbative methods. It is an analytical method which is directly based on the QCD Lagrangian.
It is a powerful tool to extract hadron properties~\cite{Shifman:1978bx, Shifman:1978by, Reinders:1984sr, Ioffe:1983ju}.
Using this method static (e.g. baryon masses) and dynamic (e.g. magnetic moments and coupling constants)
hadronic parameters can be determined.
An alternative to the traditional QCD sum rules is the
light cone QCD sum rules (LCSR). LCSR is especially suitable to study the interactions of hadrons at large momentum transfer~\cite{Braun:1988qv, Balitsky:1989ry, Chernyak:1990ag}.
In LCSR, the hadronic properties are expressed in terms of the properties of the vacuum and the light cone distribution amplitudes of  the hadrons in the process.
Since the form factors are expressed in terms of the properties of the QCD vacuum and
the distribution amplitudes, any uncertainty in these parameters reflects in the uncertainty of the predictions of the form factors.
This method has been rather successful in determining hadron form factors at high $Q^2$ (see e.g. \cite{Aliev:2004ju, Aliev:2007qu, Wang:2006su,
Braun:2006hz, Erkol:2011iw, Erkol:2011qh, Aliev:2011ku}). Using the LCSR, the isovector axial vector form factors of octet baryons have been calculated \cite{Erkol:2011qh}.
For $\Delta$ baryon isovector axial vector and pseudoscalar form factors have been studied
using  lattice QCD \cite{Alexandrou:2013opa},
chiral perturbation theory \cite{Jiang:2008we} and
quark models \cite{Choi:2010ty, Theussl:2000sj, Glozman:1997ag, Glantschnig:2004mu}.

Our aim in this work is to study the axial vector and pseudoscalar form factors in the framework of light-cone QCD sum rules for the $\Delta$ baryon.
We give the formulation of the baryon form factors on the light cone and derive our sum rules, then we present our numerical results. In the last section,
we conclude our work with a discussion on our results.

 \section{The axial and pseudoscalar form factors }

In the LCSR approach, the procedure begins with the following two-point correlation function:
\begin{equation}\label{corrf}
	\Pi_{\mu\nu}(p,q)=i\int d^4 x e^{iqx} \langle 0 |T[\eta_{\mu}^{\Delta}(0) A_\nu^{3}(x)]|\Delta(p,s)\rangle,
\end{equation}
 where $\eta_\Delta(x)$ is an interpolating current for the $\Delta$. Here we choose the interpolating current as follows \cite{Ioffe:1981kw},
  \begin{eqnarray}\label{intf}
		\eta_{\mu}^{\Delta}(0)=&\frac{1}{\sqrt{3}}\epsilon^{abc}[2(u^{aT}(0) C\gamma_\mu d^b(0)) u^c(0)+ (u^{aT}(0)C\gamma_\mu u^b(0))d^c(0)]
	\end{eqnarray}
\\
where $a$, $b$, $c$ are the color indices and $C$ denotes charge conjugation.
 The $\Delta$ matrix elements of the axial-vector transition are parametrized in terms of four invariant form factors as follows \cite{Alexandrou:2013opa};
\begin{eqnarray}
\langle \Delta(p',s')| A_\nu (x)| \Delta(p,s)\rangle = \frac {-i}{2} \overline{\upsilon}^{\alpha}(p',s')\bigg[g_{\alpha\beta}\bigg(g_1^A(q^2) \gamma_\nu \gamma_5 + g_3^A(q^2)\frac{q_\nu \gamma_5}{2M_\Delta}\bigg)  \nonumber \\
+ \frac{q^\alpha q^\beta}{4M_\Delta^{2}}\bigg( h_1^A(q^2)\gamma_\nu \gamma_5+ h_3^A(q^2)\frac{q_\nu \gamma_5}{2M_\Delta}\bigg) \bigg]\upsilon^\beta(p,s)
\end{eqnarray}
where $A^3_\nu(x) = \frac12[( \bar u(x) \gamma_\nu \gamma_5 u(x) -  \bar d(x) \gamma_\nu \gamma_5 d(x)]$ is the isovector-axial vector  current, $q = p'-p$, $M_\Delta$ is delta mass,
$\upsilon_\alpha$ is a Rarita-Schwinger spinor describing spin 3/2 fermions.
The axial charge of the $\Delta$ baryon is defined as $g^A=-3 g_1^A(0)$ where the factor $-3$ is due to the spin-3/2 nature of the $\Delta$ baryon.

In the derivation of the sum rules, summations over the spins of the $\Delta$ baryon will be performed using the summation formula for the Rarita-Schwinger spinors:
\begin{equation}\label{spinor}
\upsilon_{\mu}(p',s)\overline{\upsilon_{\nu}}(p',s)=-(\not\!p'+m_{\Delta})\{g_{\mu\nu}-\frac{1}{3}\gamma_{\mu}\gamma_{\nu}-\frac{2p'_{\mu}p'_{\nu}}
{3m_{\Delta}^{2}}+\frac{p'_{\mu}\gamma_{\nu}-p'_{\nu}\gamma_{\mu}}{3m_{\Delta}}\}
\end{equation}

Note that, isoscalar axial vector current differs from the isovector axial vector current by the relative sign of the two terms. This current has an anomaly which would contribute to the form factors. For this reason, in this work we will not study the isoscalar current.

The correlation function given in Eq. (\ref{corrf}) can be calculated  in terms of hadronic properties if $p^2>0$ and $(p+q)^2>0$, and also in terms of QCD parameters and several distribution amplitudes (DAs) of the baryon in the deep Euclidean region $p^2\rightarrow - \infty$ and $(p+q)^2 \rightarrow - \infty$.

In order to construct the sum rules for the form factors, we calculate the correlation function in terms of hadron and quark-gluon
parameters. Inserting the interpolating currents for the $\Delta$ into the correlation function in Eq. (\ref{corrf}), we determine

\begin{eqnarray} \label{eq:Qcds}
 \Pi_{\mu\nu} &=&\frac{i}{16\sqrt3}\int d^4 x e^{iqx}~(C\gamma_\mu)_{\alpha\beta}(\gamma_\nu \gamma_5)_{\rho\sigma}\left\{4\epsilon^{abc}\langle \nonumber
0|{q_1}_{\sigma}^a(0) {q_2}_{\theta}^b(x) {q_3}_{\phi}^c(0)|\Delta(p,s) \rangle \right. \\ \nonumber
& &\left.\ \bigg[2\delta_\alpha^{\eta}\delta_\sigma^{\theta} \delta_\beta^{\phi}S(-x)_{\lambda\rho}+
2\delta_\lambda^{\eta} \delta_\sigma^{\theta} \delta_\beta^{\phi}S(-x)_{\alpha\rho}\right.\nonumber
\left.+ \delta_\alpha^{\eta} \delta_{\sigma\theta} \delta_\lambda^{\phi} S(-x)_{\beta\rho}
+\delta_\beta^{\eta} \delta_\sigma^{\theta} \delta_\phi^{\lambda} S(-x)_{\alpha\rho}\bigg]\right. \nonumber \\
&&- 4\epsilon^{abc}\langle 0|{q_1}_{\sigma}^a(0) {q_2}_{\theta}^b(0) {q_3}_{\phi}^c(x)|\Delta(p,s) \rangle
\left.\bigg[ 2\delta_\alpha^{\eta}\delta_\lambda^{\theta} \delta_\sigma^{\phi} S(-x)_{\beta\rho}
+\delta_\alpha^{\eta}\delta_\beta^{\theta} \delta_\sigma^{\phi} S(-x)_{\lambda\rho}\bigg] \right\}
\end{eqnarray}
where $q_i$ ($i=1,~2,~3$)  denote the quark fields. For the case of the $\Delta$ baryon,  $q_1 = q_2 = u$ and $q_3 = d$.
Setting $m_u=m_d=0$, the light quark propagator $S(x)$ in an external gluon field can be written as:
\begin{eqnarray*}\label{qprop}
	S_q(x)=\frac{i\xslash}{2\pi^2x^4}-\frac{\langle q\bar{q}\rangle}{12}\left(1+\frac{m_0^2 x^2}{16}\right)-ig_s\int^1_0 d\upsilon\left[\frac{\xslash}{16\pi^2x^4} G_{\mu\nu}\sigma^{\mu\nu}-\upsilon x^\mu G_{\mu\nu}\gamma^\nu\ \frac{i}{4\pi^2x^2}\right].
\end{eqnarray*}
The first term of this expression describes the hard-quark propagator.
The second term  gives the contributions from the nonperturbative structure of the QCD vacuum, that is,
the quark and quark-gluon condensates. After applying Borel transformation these terms do not contribute to the sum rules. The same holds for any other contribution that contains higher order condensates which are multiplied by  positive powers of $x^2$.

The last term is due to the correction in the background gluon field and gives rise to four-particle and five-particle baryon distribution amplitudes, which are not yet known. These distribution amplitudes contain contributions from operators of higher conformal spin than the ones used in the estimation of the distribution amplitudes \cite{Braun:2000kw}. Hence, their inclusion would not be consistent with the use of the distribution amplitudes calculated in \cite{Carlson:1988gt}.
 Furthermore, in   \cite{Diehl:1998kh}, it has been argued that for the nucleon, the contribution of the higher Fock states	that will contribute to the four parton DAs is small compared to the valence quark DAs. It would be a reasonable approximation to ignore such higher Fock state contribution also for the $\Delta$ baryon.
In this work these contributions shall not be taken into account,
which leaves us with only the first term in propagator to consider.

The pseudoscalar current is defined as
\begin{eqnarray}
P(x) = \frac12\bigg( \bar u(x) \gamma_5 u(x) -  \bar d(x) \gamma_5 d(x)\bigg)
\end{eqnarray}

The  $\Delta$  matrix elements of the pseudoscalar current is decomposed in terms of two form factors as follows \cite{Alexandrou:2013opa, Alexandrou:2011py};
\begin{eqnarray}
\langle \Delta(p',s') | P(x)| \Delta(p,s)\rangle  =
- \frac{1}{2} \overline{\upsilon}_\sigma(p',s')\left[ g^{\sigma \tau}\left({\tilde g^P(q^2) }\gamma^5 \right)
+ \frac{\displaystyle q^\sigma q^\tau}
{\displaystyle 4M_\Delta^2} \left(\tilde{h^P(q^2)}\gamma^5\right) \right] \upsilon_\tau(p,s)
\end{eqnarray}

Then, inserting the interpolating current into the correlation function we determine;

\begin{eqnarray} \label{eq:Qcds1}
 \Pi_{\mu\nu} &=&\frac{i}{16\sqrt3}\int d^4 x e^{iqx}~(C\gamma_\mu)_{\alpha\beta}(\gamma_5)_{\rho\sigma}\left\{4\epsilon^{abc}\langle \nonumber
0|{q_1}_{\sigma}^a(0) {q_2}_{\theta}^b(x) {q_3}_{\phi}^c(0)|\Delta(p,s) \rangle \right. \\ \nonumber
& &\left.\ \bigg[2\delta_\alpha^{\eta}\delta_\sigma^{\theta} \delta_\beta^{\phi}S(-x)_{\lambda\rho}+
2\delta_\lambda^{\eta} \delta_\sigma^{\theta} \delta_\beta^{\phi}S(-x)_{\alpha\rho}\right.\nonumber
\left.+ \delta_\alpha^{\eta} \delta_{\sigma\theta} \delta_\lambda^{\phi} S(-x)_{\beta\rho}
+\delta_\beta^{\eta} \delta_\sigma^{\theta} \delta_\phi^{\lambda} S(-x)_{\alpha\rho}\bigg]\right. \nonumber \\
&&- 4\epsilon^{abc}\langle 0|{q_1}_{\sigma}^a(0) {q_2}_{\theta}^b(0) {q_3}_{\phi}^c(x)|\Delta(p,s) \rangle
\left.\bigg[ 2\delta_\alpha^{\eta}\delta_\lambda^{\theta} \delta_\sigma^{\phi} S(-x)_{\beta\rho}
+\delta_\alpha^{\eta}\delta_\beta^{\theta} \delta_\sigma^{\phi} S(-x)_{\lambda\rho}\bigg] \right\}
\end{eqnarray}

The matrix elements of the local three-quark operator
\begin{eqnarray*}
4\epsilon^{abc}\langle 0|q_{1\alpha}^a(a_1 x) q_{2\beta}^b(a_2 x) q_{3\gamma}^c(a_3 x)|\Delta(p,s)\rangle
\end{eqnarray*}
($a_{1,2,3}$ are real numbers denoting the coordinates of the valence quarks) can be expanded in terms of DAs using the Lorentz covariance, the spin and the parity of the baryon. Based on a conformal expansion using the approximate conformal invariance of the QCD Lagrangian up to 1-loop order, the DAs are then decomposed into local nonperturbative parameters, which can be estimated using QCD sum rules or fitted so as to reproduce experimental data. We refer the reader to Refs.~\cite{Carlson:1988gt} for a detailed analysis on DAs of $\Delta$, which we employ in our work to extract the axial-vector and pseudoscalar form factors.

The long-distance side of the correlation function is obtained using the analyticity of the correlation function, then we can write the correlation function in terms of a dispersion relation of the form
\begin{eqnarray}
\Pi_{\mu}(p,q)=\frac{1}{\pi}\int_0^\infty \frac{{Im}\Pi^\Delta_\mu(s)}{(s-p^{\prime 2})}ds
\end{eqnarray}
The ground state hadron contribution is singled out by utilizing the zero-width approximation
\begin{eqnarray}
Im\Pi_\mu=\pi \delta(s-m_{\Delta}^2)\langle 0|\eta^\Delta|\Delta(p')\rangle \langle \Delta(p')|(A_\nu(x), P(x))|\Delta(p)\rangle + \pi \rho^h(s)
\end{eqnarray}
The matrix element of the interpolating current between the vacuum and baryon state is defined as
\[\langle 0|\eta^\Delta_{\mu}|\Delta(p,s)\rangle=\lambda_\Delta\upsilon_\mu(p,s)\]
where $\lambda_\Delta$ is the baryon overlap amplitude and $\upsilon_\mu(p,s)$ is the $\Delta$ baryon spinor.

Physical part of the correlation function can be obtained as follows for axial vector current;

  \begin{eqnarray}
	\Pi_{\mu\nu} &=& i\frac {\lambda_\Delta}{2(M_\Delta^2-p'^2)} \bigg[\left\{-\frac{(3M_\Delta^{2}+4q^2)}{24M_\Delta^{5}}h_3^A(q^2)\nonumber
	-\frac{g_3^A (q^2)}{3M_\Delta^3}\right\} q_\mu q_\nu \qslash \gamma_5 q^\beta \Delta_\beta  \nonumber \\
	& &+g_1^A(q^2)\left\{-\frac{4}{3}q_\mu \gamma_5 \Delta_\nu+ p_\nu \gamma_5 \Delta_\mu+\gamma_\nu \qslash \gamma_5 \Delta_\mu- \frac{2}{3M_\Delta}q_\mu \qslash \gamma_5 \Delta_\nu\right\} -\left\{g_3^A(q^2)+ 2g_1^A(q^2)\right\} q_\nu \gamma_5\Delta_\mu \nonumber\\
	& &+\left\{\frac{(M_\Delta^{2}+2q^2)}{6M_\Delta^{4}} h_1^A(q^2)+ \frac{2g_1^A(q^2)}{3 M_\Delta^{2}}\right\}q_\mu \gamma_\nu \qslash \gamma_5 q^\beta \Delta_\beta
	+\left\{\frac{(3M_\Delta^{2}-4q^2)}{6M_\Delta^{4}}h_1^A(q^2)+\frac{2g_1^A (q^2)}{3M_\Delta^2}\right\}  q_\mu p_\nu \gamma_5 q^\beta \Delta_\beta  \nonumber \\
	& &- \frac{h_1^A(q^2)}{ M_\Delta^{3}}q_\mu p_\nu \qslash\gamma_5 q^\beta \Delta_\beta - \frac{g_3^A (q^2)}{2M_\Delta}q_\nu \qslash \gamma_5 \Delta_\mu
	+\left\{-\frac{(3M_\Delta^{2}+2q^2)}{12M_\Delta^{3}}h_1^A(q^2)\right\} q_\mu \gamma_\nu \gamma_5 q^\beta \Delta_\beta \nonumber \\	
	& &+\left\{-\frac{(-3M_\Delta^{2}+5q^2)}{12M_\Delta^{4}}h_3^A(q^2)-\frac{(5M_\Delta^{2}+2q^2)}{6M_\Delta^{4}}h_1^A(q^2)
	- \frac{4g_1^A(q^2)}{3 M_\Delta^{2}}-\frac{2g_3 (q^2)}{3M_\Delta^2} \right\}q_\mu q_\nu \gamma_5 q^\beta \Delta_\beta \bigg]
	\end{eqnarray}

for pseudoscalar current

\begin{eqnarray}
 \Pi_{\mu\nu} &=& -\frac {\lambda_\Delta}{(M_\Delta^2-p'^2)}\bigg[ \frac{\tilde g^P(q^2) M_\Delta}{6} \gamma_5\Delta_\mu  +
 \frac{\tilde h^P(q^2) }{24 M_\Delta} q_\mu \gamma_5 q^\tau \Delta_\tau \bigg]
\end{eqnarray}

In these expressions, only the contribution of the spin-$3/2$ $\Delta$ baryon is shown.
In principle, the correlation function can also include the contribution from spin-$1/2$ particles.
The overlap of the spin-$1/2$ particles with the $\eta_\mu^\Delta$ current can be written as
\begin{equation}
\langle 1/2(p') \vert \eta^\Delta_\mu \vert 0 \rangle = \left(A p'_\mu + B \gamma_\mu \right) u(p')
\end{equation}
where $u(p')$ is the spinor describing the spin-$1/2$ particle. Hence, when the gamma matrices are put into the order $\gamma_\mu\gamma_\nu\not\!q\not\!p$ in the related correlation function,
spin-$1/2$ states contribute only to the structures which have $\gamma_\mu$ at the beginning or which are proportional to $p'_\mu$.  Then the contributions of the spin-$1/2$ states in the correlation
function are eliminated by ignoring the structures proportional to
$p'_\mu$ and the structures that contain a $\gamma_\mu$ at the beginning. In this way, only
the contributions from spin-$3/2$ states are kept \cite{Belyaev:1993ss,  Belyaev:1982cd}.

The QCD sum rules are obtained by matching the short-distance calculation of the correlation
function with the long-distance calculation. As can be seen from Eq. (11), the coefficients of the structures $p_\mu \gamma_5 \Delta_\nu$ and $(q_\nu \gamma_5 \Delta_\mu-p_\mu \gamma_5 \Delta_\nu)$ are proportional to the form factors $g_1^A$ and $g_3^A$ respectively. Similarly, coefficient of the structure $\gamma_5\Delta_\mu $ in Eq. (12) is proportional to the form factor $\tilde g^P$.  Hence, in this work we will use the above mentioned structures to determine the $g_1^A$, $g_3^A$ and $g^P$ form factors. The following expressions for axial vector form factors are obtained:

\begin{eqnarray*}
	 g_1^A(q^2)  \frac{\lambda_{\Delta}}{M_\Delta-p'^{2}}&=&
	-  \frac{f_\Delta M_\Delta}{\sqrt{3}} \bigg[\int_0 ^{1}dx_2 \frac{1}{(q-px_{2})^2}\int_0^{1-x_2}d_{x_1}
	 4V(x_1,x_2,1-x_1-x_2)\nonumber \\
& &-\int_0 ^{1}dx_3 \frac{1}{(q-px_{3})^2}\int_0^{1-x_3}d_{x_1}[-T+A-2V](x_1,1-x_1-x_3,x_3)\bigg] \nonumber\\
\end{eqnarray*}
\begin{eqnarray}
    g_3^A(q^2)  \frac{\lambda_{\Delta}}{M_\Delta-p'^{2}}&=& -  \frac{f_\Delta M_\Delta}{\sqrt{3}}\bigg[\int_0 ^{1}dx_2 \frac{1}{(q-px_{2})^2}\int_0^{1-x_2}d_{x_1}
                 [2T+4A+8V](x_1,x_2,1-x_1-x_2) \nonumber \\
    & &+\int_0 ^{1}dx_3 \frac{1}{(q-px_{3})^2} \int_0^{1-x_3}d_{x_1} [-3T+3A+2V](x_1,1-x_1-x_3,x_3)\bigg] ,\nonumber\\
\end{eqnarray}

for pseudoscalar form factors \\
\begin{eqnarray}
 \tilde g^P(q^2)  \frac{\lambda_{\Delta}}{M_\Delta-p'^{2}}&=&
  \frac{6 f_\Delta M_\Delta}{\sqrt{3}} \bigg[\int_0 ^{1}dx_2 \frac{x_2^2}{(q-px_{2})^2}\int_0^{1-x_2}d_{x_1} T(x_1,x_2,1-x_1-x_2)\nonumber \\
& &+\int_0 ^{1}dx_3 \frac{x_3^2}{(q-px_{3})^2}\int_0^{1-x_3}d_{x_1}T(x_1,1-x_1-x_3,x_3)\bigg]
\end{eqnarray}

In order to eliminate the subtraction terms in the spectral representation of the correlation function,
the Borel transformation is performed.
After the transformation, contributions from excited and continuum states are also exponentially suppressed.
Then, using the quark-hadron duality
and subtracted, the contributions of the higher states and the continuum can be modeled.
Both of the Borel transformation and the subtraction of higher states
are achieved by using following substitution rules (see e.g. \cite{Braun:2006hz}):

\begin{eqnarray*}
		&\int dx \frac{\rho(x)}{(q-xp)^2}\rightarrow -\int_{x_0}^1\frac{dx}{x}\rho(x) e^{-s(x)/M^2}, \label{subtract1}\\
		\\
		&\int dx \frac{\rho(x)}{(q-xp)^4}\rightarrow \frac{1}{M^2} \int_{x_0}^1\frac{dx}{x^2}\rho(x) e^{-s(x)/M^2}+\frac{\rho(x)}{Q^2+x_0^2 m_B^2} e^{-s_0/M^2},\\
		\\
		&\int dt \frac{\rho(t)}{(q-tp)^6}\rightarrow -\frac{1}{2M^4}\int_{x_0}^1\frac{dx}{x^3}\rho(x) e^{-s(x)/M^2}-\frac{1}{2M^2}\frac{\rho(x)}{x_0(Q^2+x_0^2m_B^2)}\\
		&+\frac{1}{2}\frac{x_0^2}{Q^2+x_0^2m_B^2}\bigg[\frac{d}{dx_0}\frac{\rho(x_0)}{x_0(Q^2+x_0^2m_B^2)}\bigg]\label{subtract3}
\end{eqnarray*}
where $m_B=m_\Delta$,
\[s(x)=(1-x)m_B^2+\frac{1-x}{x}Q^2,\]
$M$ is the Borel mass  and $x_0$ is the solution of the quadratic equation for $s=s_0$:
\[x_0=\left[\sqrt{(Q^2+s_0-m_B^2)^2+4m_B^2 Q^2}-(Q^2+s_0-m_B^2)\right]/(2m_B^2),\] where $s_0$ is the continuum threshold.

\section{Numerical Analysis}
In this section, we present our numerical predictions of the axial vector and pseudoscalar form factors of $\Delta$.
To obtain the  numerical results, we need the expressions of the $\Delta$ baryon DAs which are studied in \cite{Carlson:1988gt}.
In this work, we obtain our results using the explicit expressions of DAs in Ref.\cite{Carlson:1988gt}. In order to calculate
the form factors, we need to determine the value of the residue of the $\Delta$ baryon, $\lambda_\Delta$.
We choose the value as $\lambda_\Delta= 0.038 ~GeV^3$ from analysis of the mass sum rules \cite{Aliev:2004ju, PhysRevC.57.322, Hwang:1994vp, Ioffe:1981kw},
and use the following numerical values of the parameters; $ M_\Delta = 1.23 ~GeV $~\cite{Beringer:1900zz}, $ f_\Delta = 0. 011 \pm 0.02 ~GeV^2$~\cite{Carlson:1988gt}.

In the traditional analysis of sum rules, the spectral density of
the higher states and the continuum are parametrized using quark hadron duality. In this approach,  the spectral density corresponding to the contributions of the higher states and continuum is parametrized as
$$\rho^h(s) = \rho^{QCD}(s) \theta(s - s_0).$$

The predictions for  the form factors depend on two auxiliary parameters:
the Borel mass $M^2$, and the continuum threshold $s_0$.
The continuum threshold signals the scale at which, the excited states and
continuum start to contribute to the correlation function. Hence it is expected that
$s_0\simeq(m_\Delta+0.3~GeV)^2=2.34~GeV^2$.
One approach to determine the continuum threshold and the working region of the Borel
parameter $M^2$ is to plot the dependence of the predictions on $M^2$ for a range of values
of the continuum threshold and determine the values of $s_0$ for which there is a stable
region with respect to variations of the Borel parameter $M^2$. For this reason,
in Figs.1, we plot the dependence of
the form factors $g_1^A{(Q^2)}$, $g_3^A{(Q^2)}$ and $\tilde g^P{(Q^2)}$ on $M^{2}$
for  two fixed values of $Q^2$ and for various values of $s_0$ in the range $2 ~ GeV^2 < s_0 < 4~GeV^2$.
As can be seen from these figures, for $s_0=2.5\pm0.5~GeV^2$, the predictions are practically independent of the value of $M^2$ for the shown range.
The uncertainty due to variations of $s_0$ in this range is much larger than the uncertainty due to variations with respect to $M^2$.
Note that the determined range of $s_0$ is in the range that one would expect from the physical interpretation of $s_0$.


In Figs. 2, we plot the form factors as a function of $Q^2$.
The qualitative behavior of the
form factors agree with our expectations.
The form factors $g_1^A$ and $\tilde g^P$ are the dominant axial-vector, pseudoscalar form factor
and  the only one that can be extracted directly from the matrix element at $Q^2 = 0$,
determining the axial charge and pseudoscalar form factors of
the $\Delta$.\\
 $g_1^{A(P),\Delta}$ is generally assumed to have a dipole form;

\begin{equation}
	g_1^{A(P)}(q^2) = g_1^{A(P)}(0)(1+Q^2/M_{A(P)}^2)^2
\end{equation}
Using this fit function we obtain  $M_A = 0.85~GeV$  and  $M_P= 0.74~GeV$.
Note that, in the vector meson dominance (VMD) model, the pole of the form factors is
given by the mass of the  meson that couples to the current.
The lightest axial vector meson has a mass of $m_A = 1.23$~GeV~\cite{Beringer:1900zz};
therefore our results also are smaller from the predictions of the VMD model.
We have also tried exponential form

\begin{equation}
	g_1^{A(P)}(q^2)= g_1^{A(P)}(0) \exp[-Q^2/M_{A(P)}^2]
\end{equation}
Our results are shown in Table I. The fit from $1.0~GeV^2\leq Q^2 \leq 10~GeV^2$ region produces the empirical value of $g_A$
quite well for the $\Delta$ baryon.
Besides we observe that the axial and pseudoscalar masses are very close to each other for all regions.
In Table II we present the different numerical results predicted from other theoretical models. As seen from table, our result is slightly larger than the result obtained from Lattice QCD \cite{Alexandrou:2013opa}, approximately two times smaller compared to the predictions of ChPT \cite{Jiang:2008we}  and quark models \cite{Theussl:2000sj, Glozman:1997ag, Glantschnig:2004mu}.
\begin{table}[t]
	\addtolength{\tabcolsep}{2pt}
\begin{tabular}{ccccccc}
				\hline\hline
		 Fit Region~(GeV$^2$) & $g_1^{A}(0)$ & $M_{A}$~(GeV)& \\[0.5ex]
		\hline
		$ [1.0-10]$ &1.16 & 1.15  &  \\
		$ [1.5-10]$ &0.88 & 1.24  &  \\
		$ [2.0-10]$ &0.70 & 1.32  &  \\[1ex]
\hline\hline
		 Fit Region~(GeV$^2$) & $g_{P}(0)$ & $M_{P}$~(GeV)& \\[0.5ex]
\hline\hline
		$ [1.0-10]$     &1.71 & 1.22  &  \\
	        $ [1.5-10]$ &1.50 & 1.26  &  \\
		$ [2.0-10]$     &1.31 & 1.31 &  \\
		\hline\hline
	\end{tabular}
\caption{The  values of exponential fit parameters, $g_{A(P)}$ and $M_{A(P)}$ for axial and pseudoscalar form factors.
The results include the fits from three region. }
	\label{fit_table}
\end{table}

\begin{table}[t]
	\addtolength{\tabcolsep}{2pt}
\begin{tabular}{ccccccc}
				\hline\hline
		   & \cite{Alexandrou:2013opa} &\cite{Jiang:2008we}&\cite{Theussl:2000sj}&\cite{Glozman:1997ag} &\cite{Glantschnig:2004mu}& This Work  \\[0.5ex]
		\hline
		   $g_{A}$ &$- 1.9 \pm 0.1$ & $-4.50$  &  $ - 4.47$& $- 4.48$& $- 4.30$  & $-2.70 \pm 0.6$ \\
		\hline\hline
	\end{tabular}
\caption{Different results from theoretical models which are Lattice QCD \cite{Alexandrou:2013opa}, ChPT \cite{Jiang:2008we}, quark models \cite{Theussl:2000sj, Glozman:1997ag, Glantschnig:2004mu} and also our model. }
	\label{compare}
\end{table}
\section{Conclusion}

In this study, we have calculated the isovector axial vector and pseudoscalar form factors for $\Delta$ baryon within the light cone QCD sum rules method.
We extract axial $g_1^A$, $g_3^A$ and pseudoscalar form factors $\tilde g^P$.
The form factors ($h_1^A, h_3^A $ and $ \tilde h^P$) cannot obtain from our approach since the necessary DAs have not been known yet.

Finally, we have found that our results for the $g_1^A$ and $\tilde g^P$ form factors compare well with latticeQCD,
chiral perturbation theory  and quark models results in the high-$Q^2$ region
and with those from other theoretical approaches when extrapolated to low-$Q^2$ regions via an exponential fit.
The result obtained from the axial charge exponential fit  different from other theoretical approaches.

\acknowledgments
 This work has been supported by The Scientific and Technological Research Council of Turkey (TUBITAK) under
Project No. 110T245.
 The work of A.K. and  A. O. is also partially supported by the European Union (HadronPhysics2 project Study of strongly interacting matter).

\bibliography{refs}

\begin{thebibliography}{10}

\bibitem{LopezCastro:2000cv}
G.~{Lopez Castro} and A.~Mariano.
\newblock {\em Phys.Lett.}, {\bf B 517}:339(2001).

\bibitem{Kotulla:2002cg}
M.~Kotulla, J.~Ahrens, J.R.M. Annand, R.~Beck, G.~Caselotti, et~al.
\newblock {\em Phys.Rev.Lett.}, {\bf 89}:272001, 2002.

\bibitem{Alexandrou:2013opa}
C.~Alexandrou, E.B. Gregory, T.~Korzec, G.~Koutsou, J.W. Negele, T.~Sato, and
  A.~Tsapalis.
\newblock {\em Phys.Rev.}, {\bf D87}:114513, 2013.

\bibitem{Choi:2010ty}
Ki-Seok Choi, W.~Plessas, and R.F. Wagenbrunn.
\newblock {\em Phys.Rev.}, {\bf D82}:014007, 2010.

\bibitem{Shifman:1978bx}
M.~A. Shifman, A.I. Vainshtein, and Valentin~I. Zakharov.
\newblock {\em Nucl.Phys.}, {\bf B147}:385--447, 1979.

\bibitem{Shifman:1978by}
M.~A. Shifman, A.I. Vainshtein, and Valentin~I. Zakharov.
\newblock {\em Nucl.Phys.}, {\bf B147}:448--518, 1979.

\bibitem{Reinders:1984sr}
L.J. Reinders, H.~Rubinstein, and S.~Yazaki.
\newblock {\em Phys.Rept.}, {\bf 127}:1, 1985.

\bibitem{Ioffe:1983ju}
B.L. Ioffe and Andrei~V. Smilga.
\newblock {\em Nucl.Phys.}, {\bf B232}:109, 1984.

\bibitem{Braun:1988qv}
V.~M. Braun and I.E. Filyanov.
\newblock {\em Z. Phys.}, {\bf C44}:157, 1989.

\bibitem{Balitsky:1989ry}
I.I. Balitsky, V.~M. Braun, and A.V. Kolesnichenko.
\newblock {\em Nucl. Phys.}, {\bf B312}:509--550, 1989.

\bibitem{Chernyak:1990ag}
V.L. Chernyak and I.R. Zhitnitsky.
\newblock {\em Nucl. Phys.}, {\bf B345}:137--172, 1990.

\bibitem{Aliev:2004ju}
T.M. Aliev and A.~Ozpineci.
\newblock {\em Nucl. Phys.}, {\bf B732}:291--320, 2006.

\bibitem{Aliev:2007qu}
T.M. Aliev and M.~Savci.
\newblock {\em Phys.Lett.}, {\bf B656}:56--66, 2007.

\bibitem{Wang:2006su}
Zhi-Gang Wang, Shao-Long Wan, and Wei-Min Yang.
\newblock {\em Eur.Phys.J.}, {\bf C47}:375--384, 2006.

\bibitem{Braun:2006hz}
V.M. Braun, A.~Lenz, and M.~Wittmann.
\newblock {\em Phys. Rev.}, {\bf D73}:094019, 2006.

\bibitem{Erkol:2011iw}
G.~Erkol and A.~Ozpineci.
\newblock {\em Phys.Lett.}, {\bf B704}:551--558, 2011.

\bibitem{Erkol:2011qh}
G.~Erkol and A.~Ozpineci.
\newblock {\em Phys.Rev.}, {\bf D83}:114022, 2011.

\bibitem{Aliev:2011ku}
T.M. Aliev, K.~Azizi, and M.~Savci.
\newblock {\em Phys.Rev.}, {\bf D84}:076005, 2011.

\bibitem{Jiang:2008we}
Fu-Jiun Jiang and Brian~C. Tiburzi.
\newblock {\em Phys.Rev.}, {\bf D78}:017504, 2008.

\bibitem{Theussl:2000sj}
L.~Theussl, R.F. Wagenbrunn, B.~Desplanques, and Willibald Plessas.
\newblock {\em Eur.Phys.J.}, {\bf A12}:91--101, 2001.

\bibitem{Glozman:1997ag}
L.~Ya. Glozman, Willibald Plessas, K.~Varga, and R.F. Wagenbrunn.
\newblock {\em Phys.Rev.}, {\bf D58}:094030, 1998.

\bibitem{Glantschnig:2004mu}
K.~Glantschnig, R.~Kainhofer, Willibald Plessas, B.~Sengl, and R.F. Wagenbrunn.
\newblock {\em Eur.Phys.J.}, {\bf A23}:507--515, 2005.

\bibitem{Ioffe:1981kw}
B.L. Ioffe.
\newblock {\em Nucl.Phys.}, { \bf B188}:317--341, 1981.

\bibitem{Braun:2000kw}
V.~Braun, R.J. Fries, N.~Mahnke, and E.~Stein.
\newblock {\em Nucl. Phys.}, {\bf B589}:381--409, 2000.

\bibitem{Carlson:1988gt}
Carl~E. Carlson and J.L. Poor.
\newblock {\em Phys.Rev.}, {\bf D 38}:2758, 1988.

\bibitem{Diehl:1998kh}
M.~Diehl, T.~Feldmann, R.~Jakob, and P.~Kroll.
\newblock {\em Eur. Phys. J.}, {\bf C8}:409--434, 1999.

\bibitem{Alexandrou:2011py}
C.~Alexandrou, E.~B. Gregory, T.~Korzec, G.~Koutsou, J.~W. Negele, T.~Sato, and
  A.~Tsapalis.
\newblock {\em Phys.Rev.Lett.}, {\bf107}:141601, 2011.

\bibitem{Belyaev:1993ss}
V.M. Belyaev.
\newblock {\em arXiv:hep-ph/9301257}, 1993.

\bibitem{Belyaev:1982cd}
V.M. Belyaev and B.L. Ioffe.
\newblock {\em Sov.Phys.JETP}, {\bf 57}:716--721, 1983.

\bibitem{PhysRevC.57.322}
Frank~X. Lee.
\newblock {\em Phys. Rev.}, {\bf C57}:322--328, 1998.

\bibitem{Hwang:1994vp}
W.Y.P. Hwang and Kwei-Chou Yang.
\newblock {\em Phys.Rev.}, {\bf D49}:460--465, 1994.

\bibitem{Beringer:1900zz}
J.~Beringer et~al.
\newblock {\em Phys.Rev.}, {\bf D86}:010001, 2012.

\end{thebibliography}
\newpage
\begin{figure}[htp]
\centering
 \subfloat[]{\label{fig:g1Msq.eps}\includegraphics[width=0.4\textwidth]{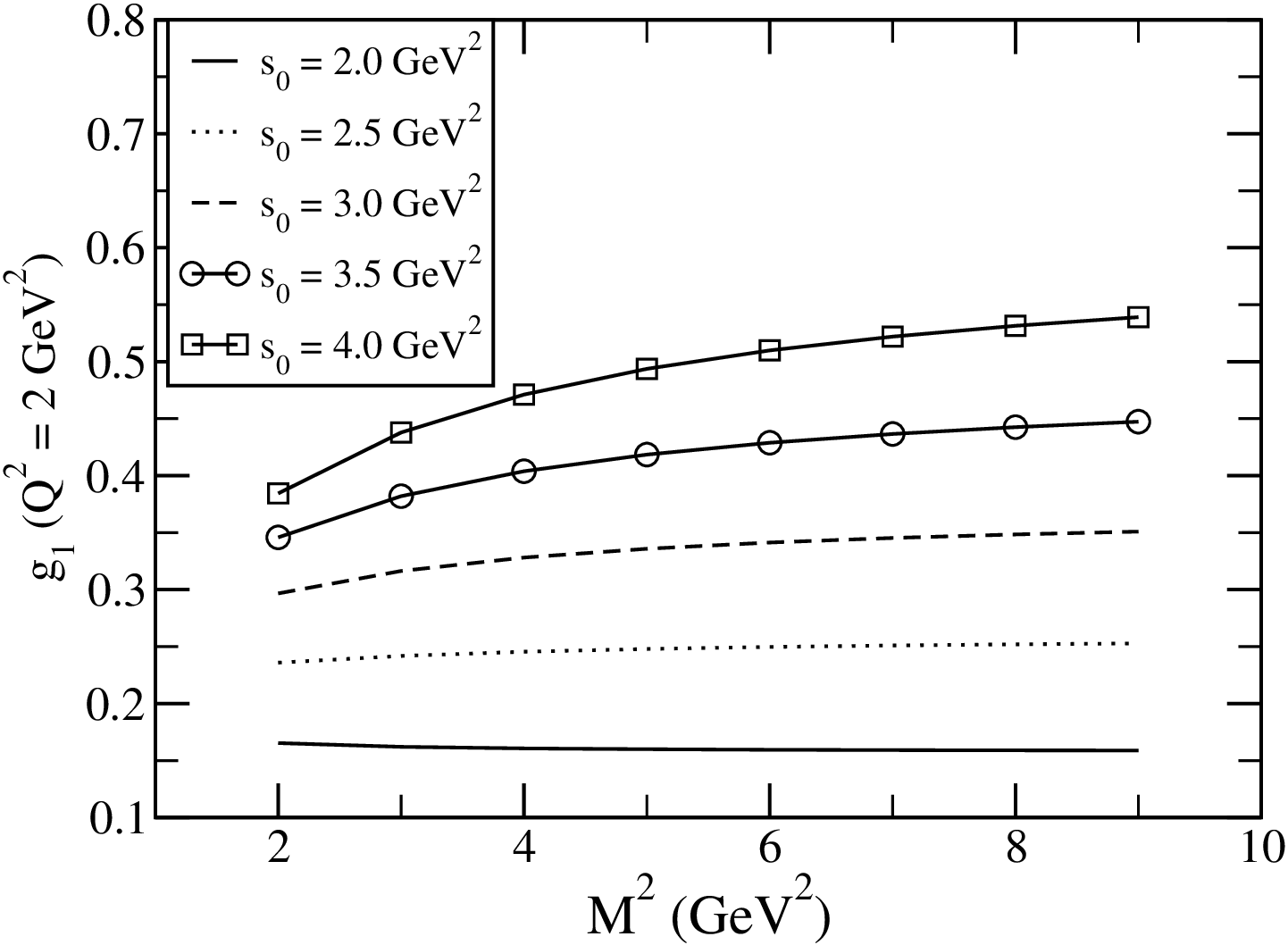}}
  \subfloat[]{\label{fig:g1Msq1.eps}\includegraphics[width=0.4\textwidth]{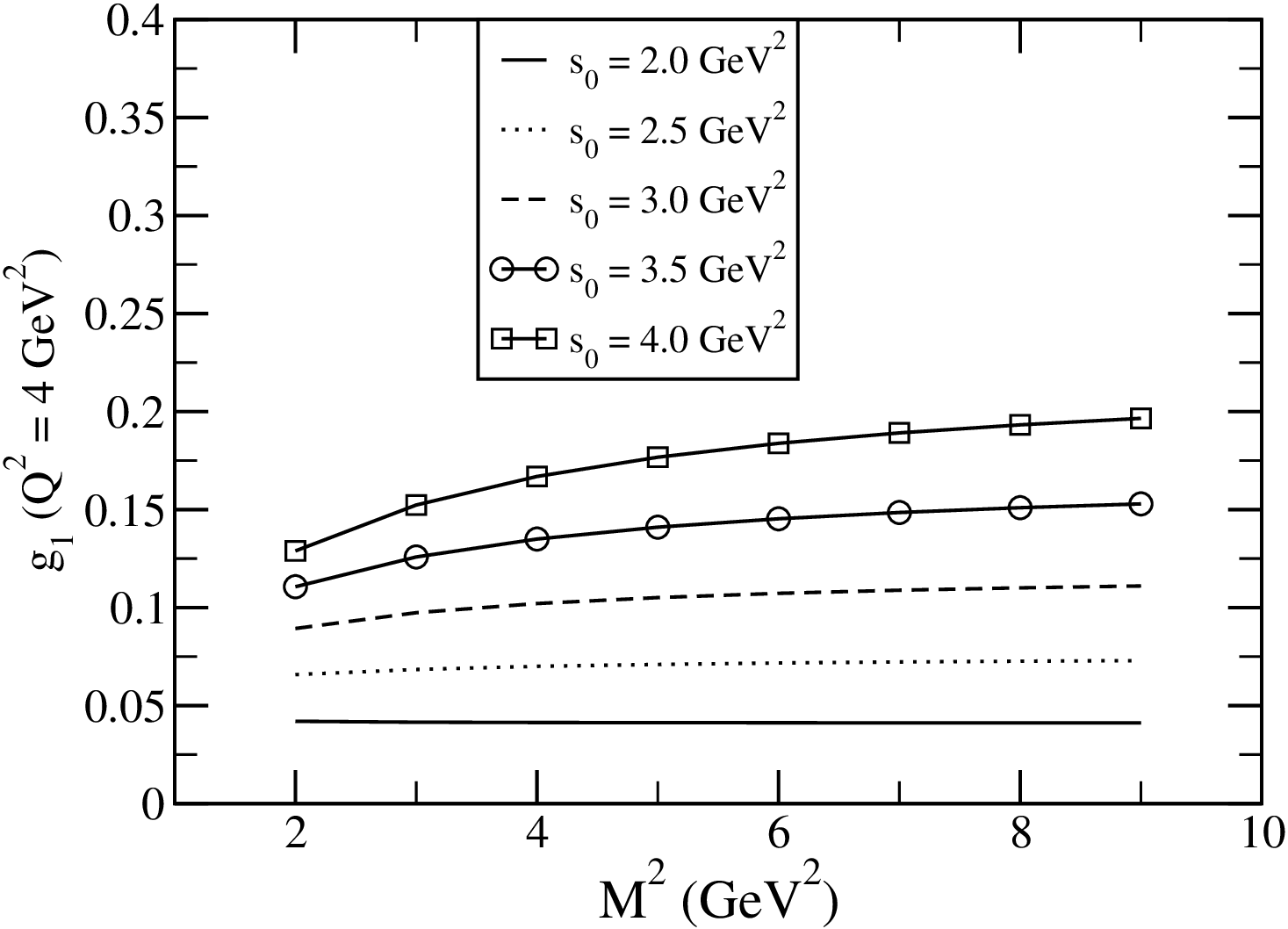}}\\
  \subfloat[]{\label{fig:g3Msq.eps}\includegraphics[width=0.4\textwidth]{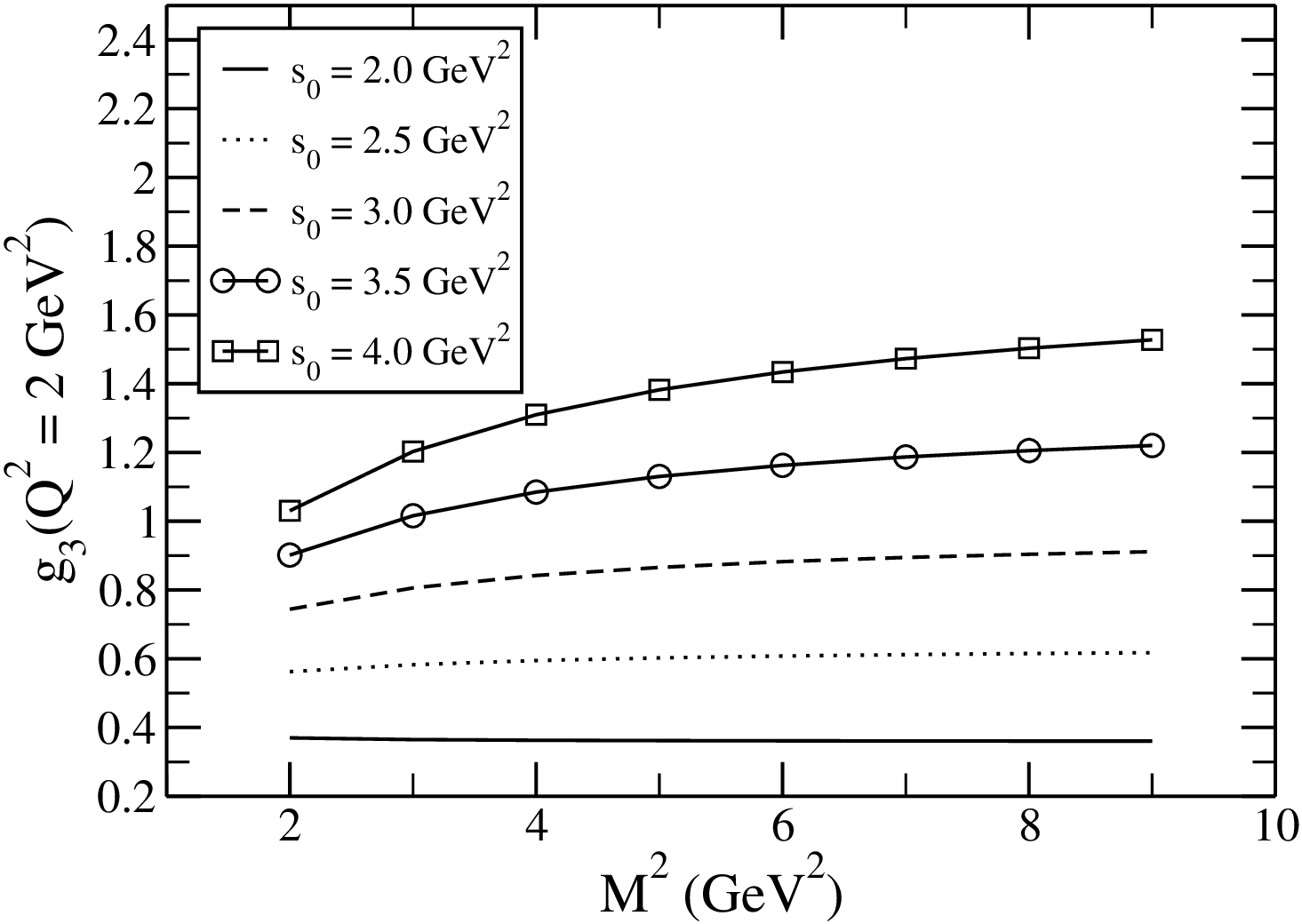}}
  \subfloat[]{\label{fig:g3Msq1.eps}\includegraphics[width=0.4\textwidth]{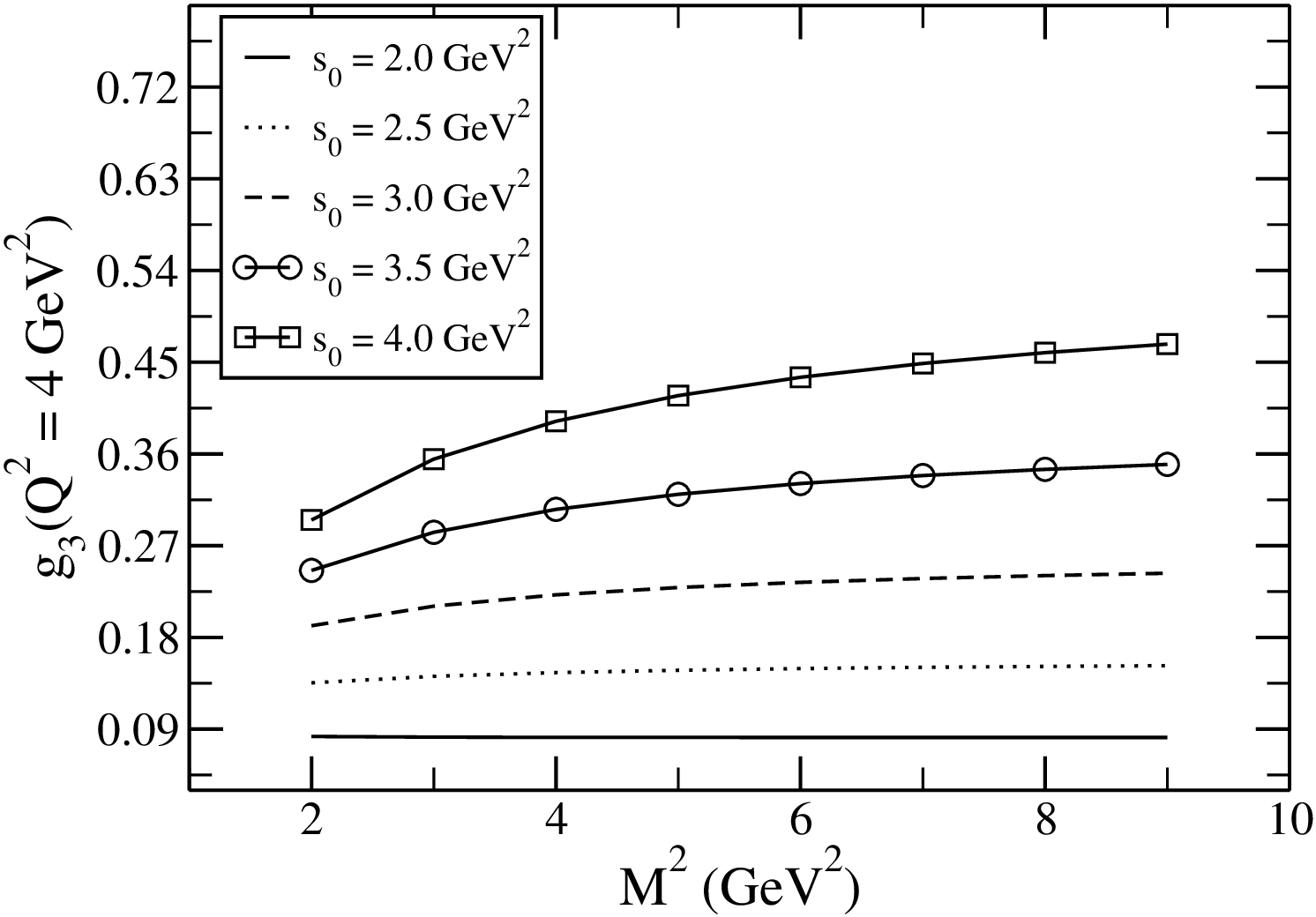}}\\
   \subfloat[]{\label{fig:gtilMsq.eps}\includegraphics[width=0.4\textwidth]{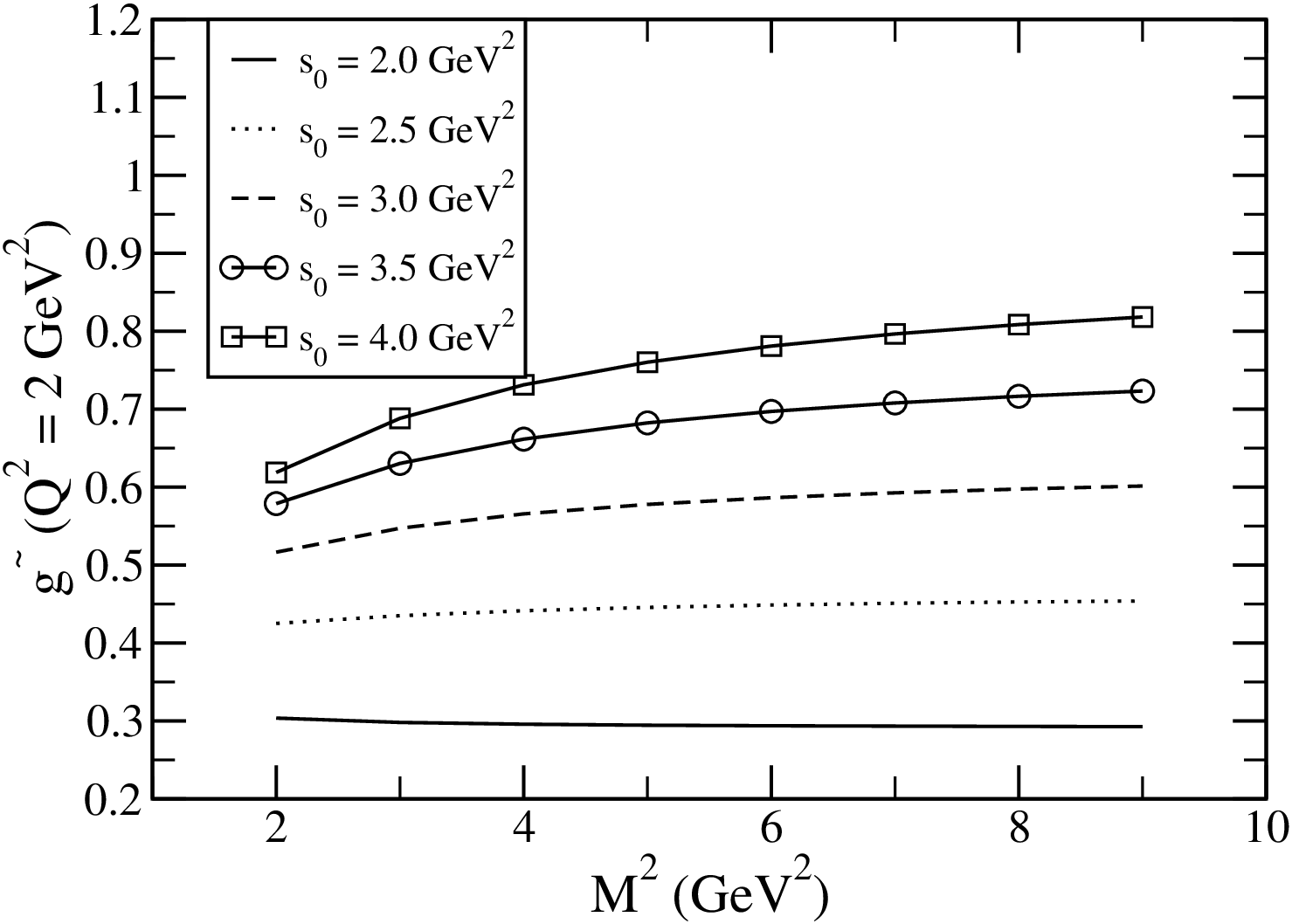}}
   \subfloat[]{\label{fig:gtilMsq1.eps}\includegraphics[width=0.4\textwidth]{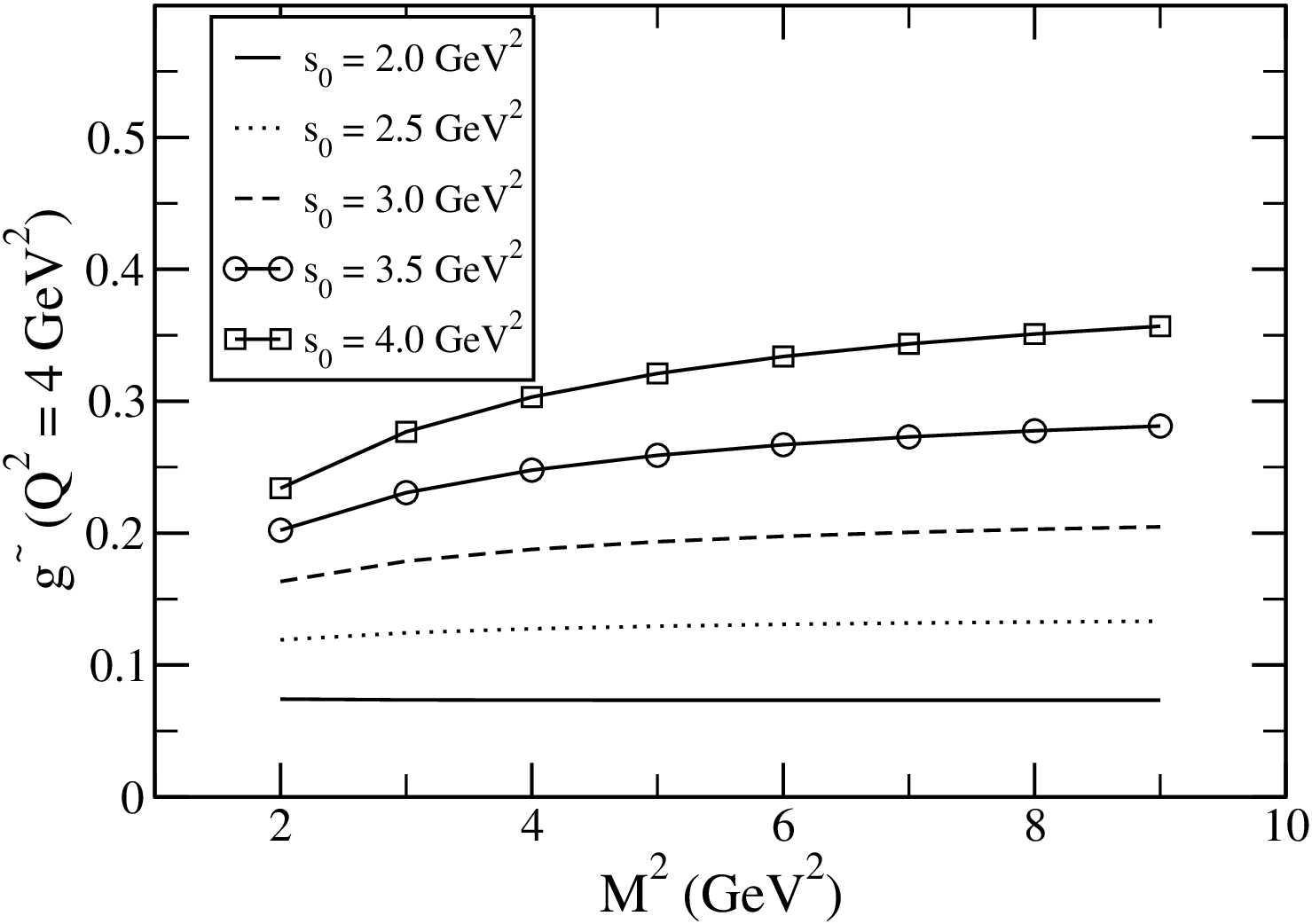}}
  \caption{ The dependence of the form factors; on the Borel parameter squared $M_B^{2}$
  for the values of the continuum threshold $s_0 = 2 ~GeV^2$, $s_0 = 2.5~GeV^2$ $s_0 = 3 ~GeV^2$,
  $s_0 = 3.5~GeV^2$ and $s_0 = 4~GeV^2$ and $Q^2 = 2$ and 4 $GeV^2$,
    (a) and (b) for $g_1^A$ axial form factor,
    (c) and (d) for $g_3^A$ axial form factor,
    (e) and (f) for $\tilde g^P(Q^2)$ pseudoscalar form factor.   }
\end{figure}

\begin{figure}[htp]
\centering
 \subfloat[]{\label{fig:g1son.eps}\includegraphics[width=0.5\textwidth]{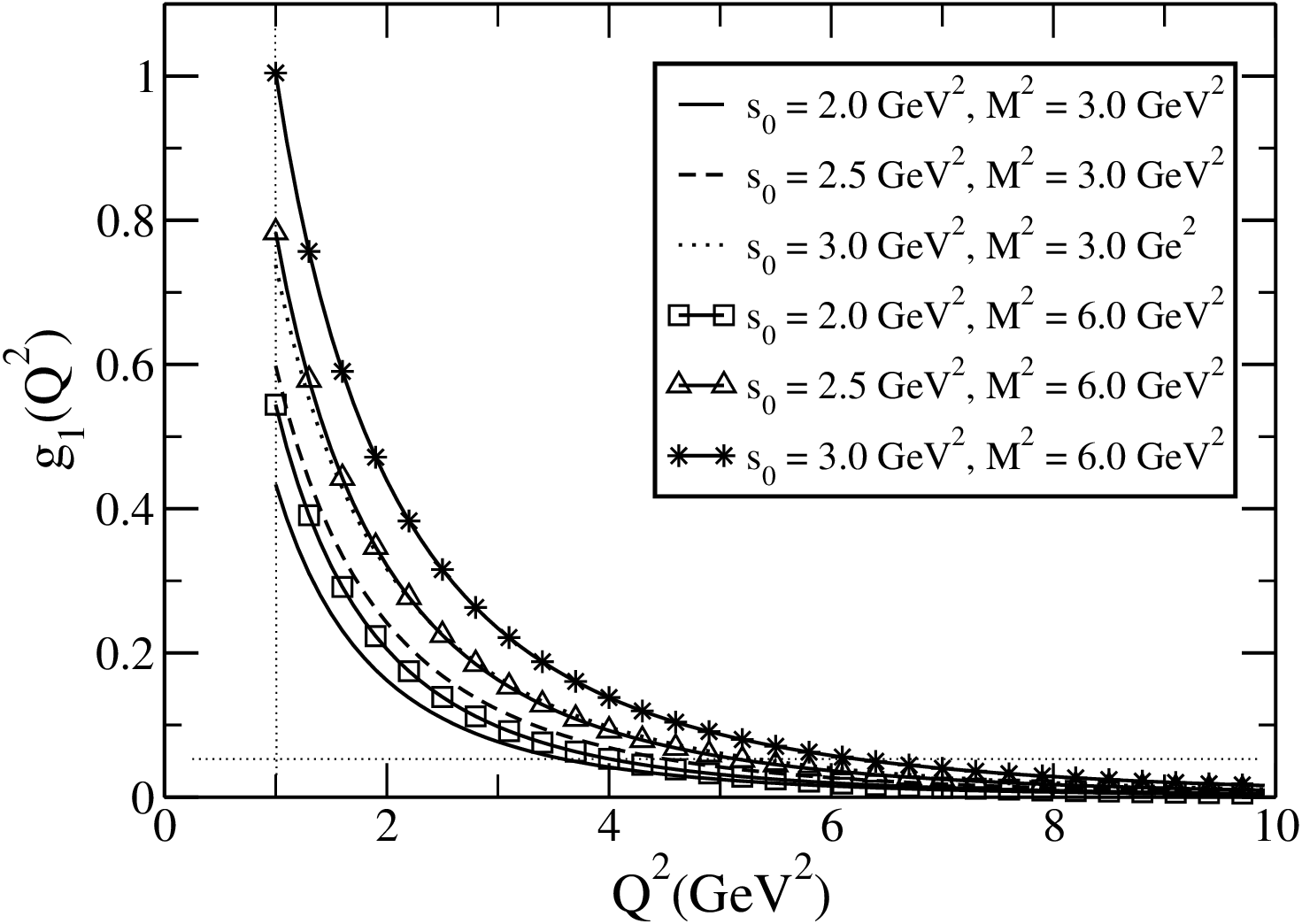}}\\
  \subfloat[]{\label{fig:g3son.eps}\includegraphics[width=0.5\textwidth]{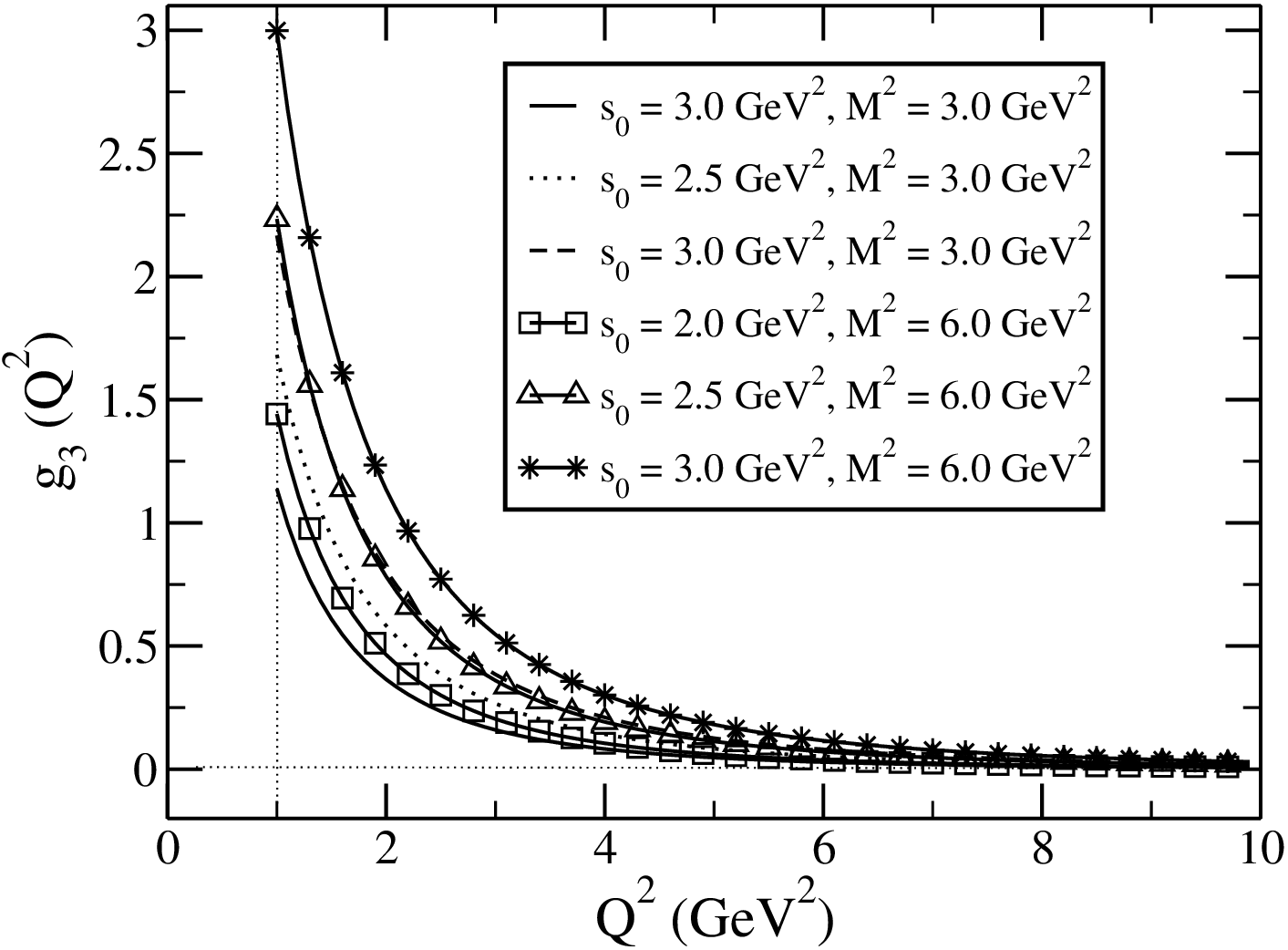}}\\
  \subfloat[]{\label{fig:gtilson.eps}\includegraphics[width=0.5\textwidth]{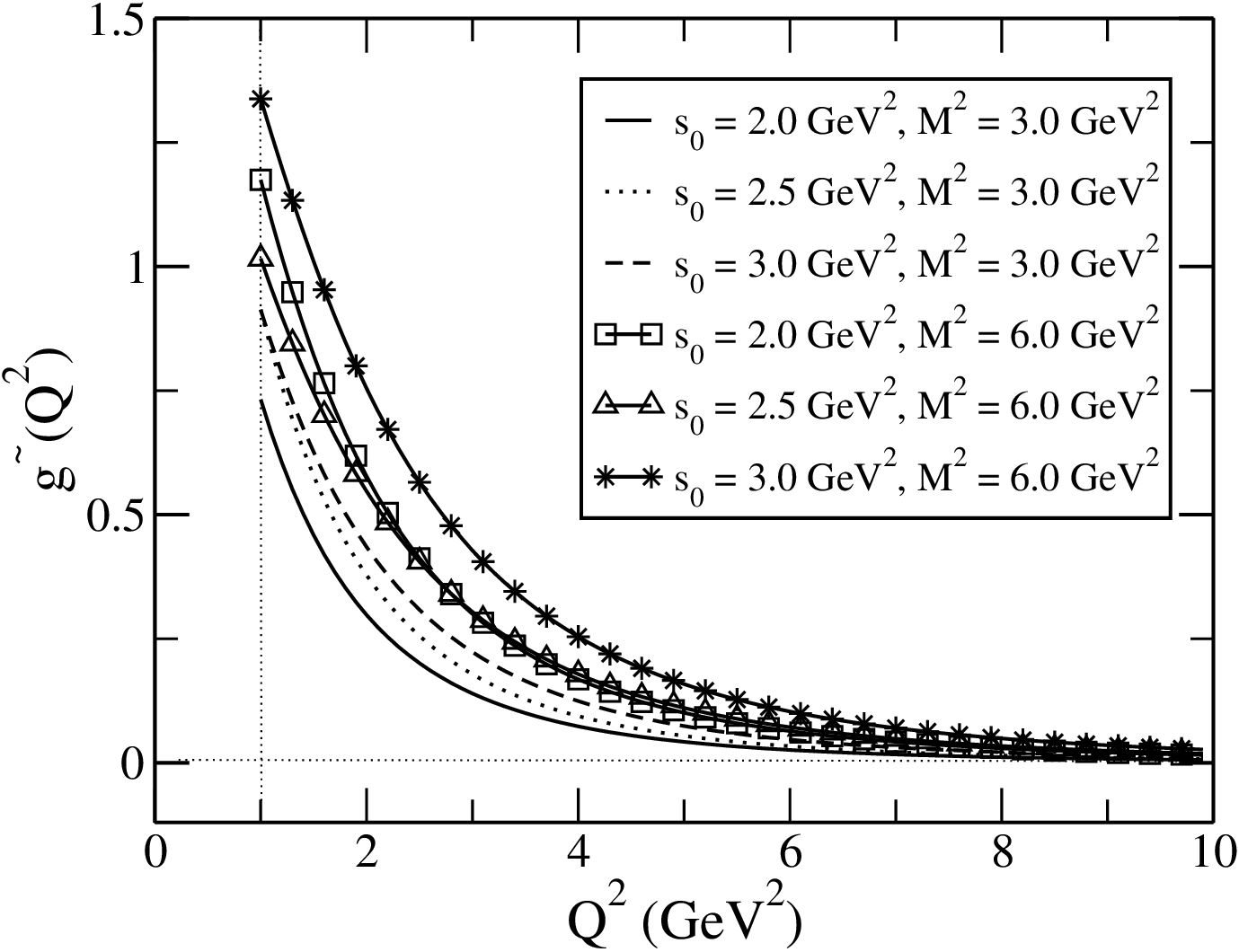}}
 \caption{The dependence of the  form factors  for the values of the continuum threshold
 $s_0 = 2 ~GeV^2$, $s_0 = 3~GeV^2$, $s_0 = 3.5~GeV^2$ and $M^{2}=3$ and $6~GeV^2$
    (a) for $g_1^A$ axial form factors,
    (b) for $g_3^A$ axial form factor,
    (c) for $\tilde g^P(Q^2)$ pseudoscalar form factor.}
\end{figure}
\end{document}